\documentclass[preprint,showpacs,preprintnumbers,
               superscriptaddress,floatfix]{revtex4}

\usepackage{amssymb}
\usepackage{graphicx}
\usepackage{dcolumn}
\usepackage{bm}
\usepackage{longtable}

\begin{document}

\title{Spin Symmetry for Anti-Lambda Spectrum in atomic nucleus\footnote{Supported by the National Key Basic Research Programme of
China under Grant No 2007CB815000, the National Natural Science
Foundation of China under Grant Nos 10775004, the Southwest
University Initial Research Foundation Grant to Doctor (No.
SWU109011). Helpful discussions with Avraham Gal are gratefully
acknowledged.}}

\author{SONG Chun-Yan}
\affiliation{School of Phyics, and State Key Laboratory of Nuclear
Physics and Technology, Peking University, Beijing, 100871}
\author{YAO Jiang-Ming}
\affiliation{School of Phyics, and State Key Laboratory of Nuclear
Physics and Technology, Peking University, Beijing, 100871}
\affiliation{School of Physics, Southwest University, Chongqing,
400715}
\author{MENG Jie\footnote{Email: mengj@pku.edu.cn}}
\affiliation{School of Phyics, and State Key Laboratory of Nuclear
Physics and Technology, Peking University, Beijing, 100871}
\affiliation{Center of Theoretical Nuclear Physics, National
Laboratory of Heavy Ion Accelerator, Lanzhou 730000}
\affiliation{Institute of Theoretical Physics, Chinese Academy of
Sciences, Beijing 100080} \affiliation{Department of Physics,
University of Stellenbosch, Stellenbosch, South Africa}
\date{\today}

\begin{abstract}
The spin symmetry of anti-Lambda spectrum in nucleus $^{16}$O has
been studied in the relativistic mean field theory. The spin-orbit
splittings of spin doublets are found to be around 0.03-0.07 MeV and
the dominant components of the Dirac spinor for the anti-Lambda spin
doublets are found to be near identical. It indicates that there is
an even better spin symmetry in the anti-Lambda spectrum than that
in the anti-nucleon spectrum.

\end{abstract}

\maketitle

 Symmetries in single particle spectrum of atomic nuclei have
been discussed extensively in the literature, as the violation of
spin-symmetry by the spin-orbit term and approximate pseudo-spin
symmetry in nuclear single particle spectrum: atomic nuclei are
characterized by a very large spin-orbit splitting, i.e., pairs of
single particle states with opposite spin ($j=l\pm
\displaystyle\frac{1}{2}$) have very different energies\cite{MY.55}.
This fact has allowed the understanding of magic numbers in nuclei
and forms the basis of nuclear shell structure. More than thirty
years ago
 pseudo-spin quantum numbers have been introduced
by $\tilde{l}=l\pm 1$ and $\tilde{j}=j$ for $j=l\pm
\displaystyle\frac{1}{2}$ and it has been observed that the
splitting between pseudo-spin doublets in nuclear single particle
spectrum is by an order of magnitude smaller than the normal
spin-orbit splitting\cite{AHS.69,HA.69}.

The relativistic mean field (RMF) theory has been widely used for
describing nuclear matter, finite nuclei and
hypernuclei\cite{MengPPNP06}. Since the relation between the
pseudospin symmetry and the RMF theory was first noted in
Ref.\cite{Bahri.92}, the RMF theory has been extensively used to
describe the pseudospin symmetry in the nucleon spectrum. In
Refs.~\cite{Blokhin.95}, it suggested that the origin of pseudospin
symmetry is related to the strength of the scalar and vector
potentials. Ginocchio took a step further to reveal that
pseudo-orbital angular momentum is nothing but the ``orbital angular
momentum" of the lower component of the Dirac wave function, and
showed clearly that the origin of pseudo-spin symmetry in nuclei is
given by a relativistic symmetry in the Dirac
Hamiltonian\cite{Gin.97}. The quality of pseudo-spin symmetry has
been found to be related to the competition between the centrifugal
barrier and the pseudo-spin orbital potential \cite{MSY.98,MSY.99}
with the RMF theory.

The possibility of producing a new nuclear system with one or more
anti-baryons inside normal nuclei has recently gained renewed
interest \cite{BMS.02,Mishustin05,Friedman05,Larionov08,Larionov09}.
In Ref.\cite{Zhou03}, the RMF theory has been used to investigate
the antinucleon spectrum, which corresponds to the negative energy
solutions to the Dirac equation, and a well developed spin symmetry
has been found in the antinucleon spectrum. It motivates the present
study of the spin symmetry in the single $\bar{\Lambda}$ spectrum,
which can provide us important information on the antiparticles and
their interaction with nuclei.

In the RMF theory, the $\bar{\Lambda}$ hyperon is described as a
Dirac spinor moving in the potentials generated by the meson fields,
 \begin{equation}
 \left\{ \bm{\alpha}\cdot \bm{p}
 + V_{\bar{\Lambda}}(\bm{r})
 + \beta [M_{\bar{\Lambda}}
 +S_{\bar{\Lambda}}(\bm{r})]
 \right\} \psi_{\bar{\Lambda}}(\bm{r})
 = \epsilon_{\bar{\Lambda}} \psi_{\bar{\Lambda}}(\bm{r}),
 \label{eq:Dirac0}
\end{equation}
where $M_{\bar{\Lambda}}$ is the mass of $\bar\Lambda$ and chosen as
$M_{\bar{\Lambda}}=1115.7$ MeV, $\epsilon_{\bar{\Lambda}}$ is the
single-particle energy. As $\bar{\Lambda}$ is a charge neutral and
isoscalar particle, it couples only to the $\sigma$ and $\omega$
mesons. As a consequence, the scalar $S_{\bar{\Lambda}}(\bm{r})$ and
vector $V_{\bar{\Lambda}}(\bm{r})$ potentials are given by,
\begin{eqnarray}
S_{\bar{\Lambda}}(\bm{r})=g_{\sigma \bar{\Lambda}}\sigma(\bm{r}),\\
V_{\bar{\Lambda}}(\bm{r})=g_{\omega \bar{\Lambda}}\omega(\bm{r}).
\end{eqnarray}
 According to the charge conjugation transformation, the coupling constants for $\bar\Lambda$
 are related to those for $\Lambda$ by the following relations,
\begin{eqnarray}
g_{\sigma \bar{\Lambda}}&=&g_{\sigma \Lambda}\\
g_{\omega \bar{\Lambda}}&=&-g_{\omega \Lambda}.
\end{eqnarray}

For a spherical system, the Dirac spinor of $\bar{\Lambda}$ is
characterized with quantum numbers $\{nl\kappa m\}$ and has the form
\begin{equation}
 \psi_{\bar{\Lambda}}(\bm{r}) = \displaystyle\frac{1}{r}
  \left(
   \begin{array}{c}
    i G_{n\kappa}(r)         Y_{jm}^{l}(\theta,\phi)        \\
    - F_{\tilde{n}\kappa}(r) Y_{jm}^{\tilde{l}}(\theta,\phi)
   \end{array}
  \right) ,
  \ \ j=l\pm \displaystyle\frac{1}{2},
 \label{eq:SRHspinor}
\end{equation}
where $Y_{jm}^{l}(\theta,\phi)$ are the spin spherical harmonics,
$G_{n\kappa }(r)/r$ and $F_{\tilde{n}\kappa }(r)/r$ form the radial
wave functions for the upper and lower components with $n$ and
$\tilde{n}$ radial nodes, and $\kappa = \langle 1 +
\mathbf{\bm{\sigma} \cdot l} \rangle = (-1)^{j+l+1/2}(j+1/2)$
characterizes the spin orbit operator and the quantum numbers $l$
and $j$.

Using the relation between the upper and lower components, one can
obtain the Schr\"{o}dinger-like equations for the upper (dominant)
 component from Eq.(\ref{eq:Dirac0}),
 \begin{widetext}
 \begin{equation}
 \label{eq:origin-of-symmetry1}
 \left[
  - \displaystyle\frac{1}{2M_{+}}
  \left( \displaystyle\frac{d^{2}}{dr^{2}}
        +\displaystyle\frac{1}{2M_{+}}\displaystyle\frac{dV_{-}}{dr}\displaystyle\frac{d}{dr}
        -\displaystyle\frac{l(l+1)}{r^{2}}
  \right)
  - \displaystyle\frac{1}{4M_{+}^{2}} \frac{\kappa}{r} \displaystyle\frac{dV_{-}}{dr}
  + M_{\bar{\Lambda}}-V_{+}
 \right] G(r)
 = \epsilon_{\bar{\Lambda}} G(r),
 \end{equation}
\end{widetext}
where $M_{+} = M_{\bar{\Lambda}} + \epsilon_{\bar{\Lambda}} - V_{-}$
and $V_{\pm}(r) = V_{\bar{\Lambda}}(r)\pm S_{\bar{\Lambda}}(r)$. In
Eq.~(\ref{eq:origin-of-symmetry1}), it is shown that the spin
symmetry depends on the competition between the spin-orbit term (the
term $\sim \kappa$) and the centrifugal term~\cite{MSY.98}. In the
following, the spin symmetry in $\bar\Lambda$ spectrum in atomic
nucleus will be studied by taking $^{16}$O as an example.

With the mean-field and no-sea approximations, the coupled Dirac
equations for nucleons and $\bar{\Lambda}$ together with the
Klein-Gordon equations for mesons can be self-consistently solved by
iteration. The effective interaction PK1~\cite{Long04} is adopted
for the nucleon part, and the coupling constants for $\bar{\Lambda}$
are chosen as, $g_{\sigma \bar{\Lambda}}=g_{\sigma
\Lambda}=\displaystyle\frac{2}{3}g_{\sigma N}$ and $g_{\omega
\bar{\Lambda}}=-g_{\omega
\Lambda}=-\displaystyle\frac{2}{3}g_{\omega N}$, according to the
SU(3) symmetry in naive quark model.

\begin{figure}[tbp]
\includegraphics[width=8.2cm]{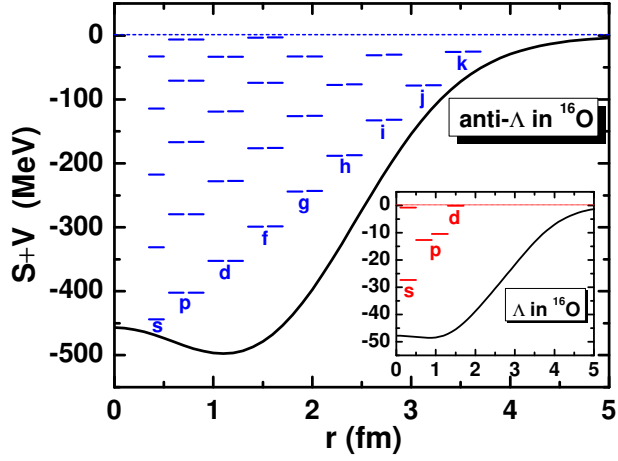}
\caption{Potential and spectrum of $\bar{\Lambda}$ in $^{16}$O. For
each pair of the spin doublets, the left levels are with $\kappa<0$
and the right ones with $\kappa>0$. The inset gives the potential
and spectrum of $\Lambda$ in $^{16}$O.} \label{fig:spec-o16-AL}
\end{figure}

The potential and single $\bar{\Lambda}$ spectrum in $^{16}$O are
plotted in Fig.~\ref{fig:spec-o16-AL}, where for each pair of the
spin doublets, the left levels are with $\kappa<0$ and the right
ones with $\kappa>0$. For comparison, the potential and single
$\Lambda$ spectrum in $^{16}$O are given as well. As seen in
Fig.~\ref{fig:spec-o16-AL}, the single anti-Lambda energies for each
spin doublets are almost identical, and the energy differences
between spin doublets
$\epsilon_{\bar{\Lambda}(nl_{j=l-1/2})}-\epsilon_{\bar{\Lambda}(nl_{j=l+1/2})}$
in the anti-Lambda spectrum are around 0.09-0.17 MeV for p states,
which are much smaller than that in Lambda spectrum, 2.26 MeV.

\begin{figure}[tbp]
\includegraphics[width=8.2cm]{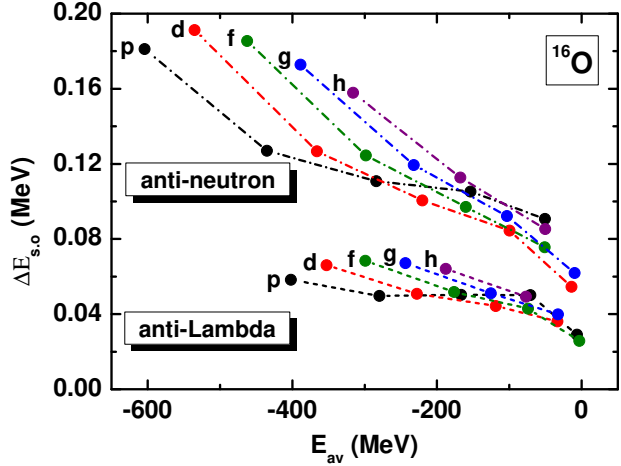}
\caption{The spin-orbit splitting $\Delta E_{s.o}$ [defined in
Eq.(\ref{sosplitting})] for $\bar{\Lambda}$ and antineutron in
$^{16}$O as a function of the average energy $E_{\rm av}$ [defined
in Eq.(\ref{average})]. For each orbit $l$, the radial quantum
numbers $n$ of the spin doublets are $1,2,\cdots$, respectively from
left to right.} \label{fig:spli-AL}
\end{figure}

In order to see the splitting size and its energy dependence more
clearly, the $\bar{\Lambda}$ spin-orbit splitting,
 \begin{equation}
 \label{sosplitting}
 \Delta E_{s.o}=(\epsilon_{\bar{\Lambda}(nl_{j=l-1/2})}-\epsilon_{\bar{\Lambda}(nl_{j=l+1/2})})/(2l+1),
 \end{equation}
in $^{16}$O is plotted as a function of the the average energy,
 \begin{equation}
 \label{average}
  E_{\rm av}=\displaystyle\frac{1}{2}(E_{\bar{\Lambda}(nl_{j=l-1/2})}+E_{\bar{\Lambda}(nl_{j=l+1/2})})
 \end{equation}
in Fig.~\ref{fig:spli-AL}, where
$E_{\bar{\Lambda}}=\epsilon_{\bar{\Lambda}}-M_{\bar{\Lambda}}$. For
comparison, the spin-orbit splittings for antineutron are also
plotted here. It is seen that the spin-orbit splittings $\Delta
E_{s.o}$ for p states in anti-Lambda spectrum are around 0.03-0.06
MeV, which is much smaller than those both in Lambda spectrum, 0.75
MeV, and in antineutron spectrum, 0.09-0.18 MeV. It indicates that
an even better spin symmetry in $\bar{\Lambda}$ spectrum has been
found than that in antineutron spectrum~\cite{Zhou03}. This can be
easily understood from
 $V^{\bar\Lambda}_-(r)=\displaystyle\frac{2}{3}V^{\bar n}_-(r)$,
 and
 $M^{\bar \Lambda}_+ > M^{\bar n}_+$
 due to
 $M^{\bar \Lambda}_+=M_{\bar{\Lambda}} -(- \epsilon_{\bar{\Lambda}}
 + V^{\bar\Lambda}_-)$,
 $M^{\bar n}_+ =M_{\bar{n}}-(-\epsilon_{\bar{n}}+ V^{\bar n}_-)$, and
 $M_{\bar{\Lambda}}>M_{\bar{n}}$,
$- \epsilon_{\bar{\Lambda}}<-\epsilon_{\bar{n}}$,
 where $M_{\bar n}$ is the mass of anti-neutron.
Therefore the spin-orbit term $\displaystyle\frac{1}{4M_{+}^{2}}
\displaystyle\frac{\kappa}{r} \displaystyle\frac{dV_{-}}{dr}$ in the
Eq.~(\ref{eq:origin-of-symmetry1}) for $\bar{\Lambda}$ is smaller
than two thirds of that for antineutron, which leads to an even
better spin symmetry in $\bar{\Lambda}$ spectrum than that in
antineutron ones. Moreover, it is found that the spin-orbit
splittings for other orbits are around 0.03-0.07 MeV, which
decreases as the levels approaching to the continuum.

\begin{figure}[tbp]
\includegraphics[width=8.0cm]{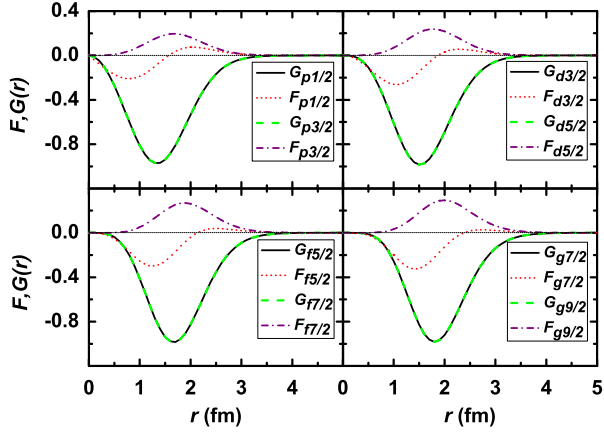}
\caption{Radial wave functions of $\bar{\Lambda}$ spin doublets with
different orbit quantum numbers in $^{16}$O.}
\label{fig:wf-o16-pdfg}
\end{figure}

In Fig.~\ref{fig:wf-o16-pdfg}, we plot the radial wave functions
$G(r)$ and $F(r)$ for several $\bar{\Lambda}$ spin doublets in
$^{16}$O with different orbit quantum numbers. Since the spin-orbit
splitting in the single $\bar{\Lambda}$ spectrum in $^{16}$O is so
small, the dominant component $G(r)$ of the wave functions of the
spin doublets are nearly identical while the small components $F(r)$
are quite different. The relation between the node numbers of large
and small components for the spin-doublets~\cite{LG.01,Zhou03} can
be clearly seen in Fig.~\ref{fig:wf-o16-pdfg}, i.e.,
\begin{equation}
 \tilde{n}  = n+1,\ \text{for}\ \kappa>0;
 \ \tilde{n}= n,\ \text{for}\ \kappa<0.
 \label{eq:node2}
\end{equation}

In summary, taking $^{16}$O as an example, the spin symmetry in the
single $\bar{\Lambda}$ spectrum has been studied within the RMF
theory. The spin-orbit splittings in the $\bar{\Lambda}$ spectrum
have been found to be around 0.03-0.07 MeV, which are much smaller
than those in antineutron spectrum 0.06-0.20 MeV. The reasons are
due to the shallower potential and larger mass for $\bar{\Lambda}$
hyperon. The dominant components of the Dirac spinor for the
$\bar{\Lambda}$ spin doublets are found to be near identical. All
these show that an even better spin symmetry than that in the
anti-nucleon spectrum~\cite{Zhou03} can be expected in the
anti-Lambda spectrum.

It has to be pointed out that, as in the anti-nucleon
case~\cite{Zhou03}, the imaginary part of the optical potential of
anti-Lambda has not been taken into account here, which is large and
may make it difficult to observe the spin-orbit splitting of the
anti-Lambda levels experimentally. In addition, the polarization
effects due to the $\bar{\Lambda}$ is neglected here. For a real
$\bar{\Lambda}$-$^{16}\mathrm{O}$ ($^{17}_{\bar{\Lambda}}$O) system,
the mean fields including the scalar and vector ones will be
modified by the added $\bar{\Lambda}$~\cite{BMS.02}. The study of
spin symmetry in anti-Lambda spectrum with these self-consistences
is in progress.

It is known that the tensor coupling will cancel with the original
spin-orbit potential and give a small the spin-orbit splitting in
Lambda hypernucleus~ \cite{Jennings90, Noble80, Yao08}. For
anti-Lambda hypernuclei, although the tensor coupling is expected to
increase the spin-orbit splitting, the conclusion that spin symmetry
in anti-Lambda is better than that in anti-neutron will remain true
as the contribution of the tensor couplings is comparable in
magnitude with the original spin-orbit potential for
$\bar{\Lambda}$. The detailed quantitative studies of this effect
will be given elsewhere.

\end{document}